# Photoemission study on pristine and Ni-doped SrTiO$_3$ thin films

Fatima Alarab[1,2,3], Karol Hricovini[1,3], Berengar Leikert[4], Laurent Nicolai[2], Mauro Fanciulli[1,3], Olivier Heckmann[1,3], Christine Richter[1,3], Lucie Prušakova[2], Zdeňek Jansa[2], Pavol Šutta[2], Julien Rault[5], Partick Lefevre[5], Matthias Muntwiller[6], Ralph Claessen[4], and Ján Minár[2]

[1]LPMS, CY Cergy Paris Université, Neuville-sur-Oise, France
[2]New Technologies Research Centre, University of West Bohemia, Pilsen, Czech Republic
[3]LYDIL, CEA, Université Paris Saclay, Gif-sur-Yvette, France
[4]Physikalisches Institut und Röntgen Center for Complex Materials (RCCM), Universität Würzburg, Germany
[5]Synchrotron SOLEIL, Saint-Aubin, Gif-sur-Yvette, France and
[6]PSI, Villigen, Switzerland

We combined photoemission spectroscopy with first-principle calculations to investigate structural and electronic properties of SrTiO$_3$ doped with Ni impurities. In SrTiO$_3$ polycrystalline thin films, grown by magnetron sputtering, the mean size of the crystallites increases with the concentration of Ni. To determine the electronic band structure of SrTiO$_3$ films doped with Ni, high quality ordered pristine and SrTiO$_3$:Ni$_x$ films with x=0.06 and 0.12 were prepared by pulsed laser deposition. Electronic band structure calculations for the ground state, as well as one-step model photoemission calculations, which were obtained by means of the Korringa-Khon-Rostoker Greens's function method, predicted the formation of localised 3d-impurity bands in the band gap of SrTiO$_3$ close to the valence band maxima. The measured valence bands at the resonance Ni2p excitation and band dispersion confirm these findings.

**I. INTRODUCTION**

Transition metal oxides in general, and strontium titanate (SrTiO$_3$/STO) in particular, with perovskite structure, are receiving a large attention from scientists because of their outstanding properties, such as high temperature superconductivity [1-3], large negative magnetoresistance [4], multiferroicity [5] and the formation of a two-dimensional electron gas (2DEG) at oxide interfaces [6-9].

They have been studied in recent years, as flexible and adaptable materials for several technological applications, especially for harvesting solar light and for photo-catalysis [10-18]. STO is active only under UV light and remains transparent for visible light due to its wide band gap which is about 3.2 eV at room temperature [19]. This inactivity in the visible light calls to solve the problem by band gap tuning. In order to adjust the optical absorption spectrum one can manipulate the composition of the material through chemical doping or alloying. Theoretical and experimental studies have pointed out that these lattice defects from different types of atoms and doping sites are not only able to adjust the absorption edge, as for instance, in BaTiO$_3$, Fe-doping can extend the absorption edge to around 647 nm [20], but also modify in different ways the electronic structure and electronic conductivity of these materials. In fact, transition metal (TM)-doping of perovskite oxides is one of the most effective strategies to absorb visible light due to the 3d-bands of the dopant which can cause a shift of the valence band and/or the conduction band and form new energy levels in the band gap [21-23].

Cations are in general implemented into the perovskite lattice by substitution with charge compensation in a number of ways. According to the defect formation energies calculated in previous studies, TM elements are more likely favorable to replace the Ti-site leaving defects with an effective negative charge relative to the host lattice [22, 24]. Recent studies on STO doped with several 3d-impurities have shown that Ni is one of the most promising dopants from the point of view of the effective sunlight absorption [25]. In fact, the substitution of Ti atoms at the B site in STO with nickel (Ni), a divalent impurity with d$^8$ electronic configuration, compensated by oxygen vacancy reduces the band gap. The valency and chemical environment of Ni impurities in STO have been studied by many authors [25-31]. Some of the data indicated that Ni ions are present in the form of Ni$^{2+}$ in cubic sites and are substitutional on Ti$^{4+}$ sites [30], while others attested the presence of features related to Ni$^{3+}$ with one oxygen vacancy [29] and even close to Ni$^{4+}$ [25]. The absorption edge shift of STO:Ni with respect to NiO was found to be 1.1 eV in [31] and 2.5 eV in [25]. Dissimilarity of the observed features was also present in the sample colour. It was reported to be beige by [31] and black by [25].

Experimental data on the electronic structure of STO:Ni and the formation of Ni impurity bands are rare, yet several theoretical investigations from first-principles calculations are available [22, 25, 32]. Nevertheless, the availability of a variety of experimental and theoretical results made it difficult to undertake a direct comparison due to the use of different methods for sample preparation and associated characterization techniques [24, 33, 34].

In this paper, we first present an experimental study of STO:$Ni_x$ ($x_{Ni}$ = 4,7%, 6% and 8%) films deposited by magnetron sputtering. In the second part, a detailed study of the electronic structure of STO:$Ni_x$ (100) thin films with ($x_{Ni}$ = 6% and 12%) deposited by pulsed laser deposition (PLD) is presented. The experimental results are compared to DFT calculations based on the KKR-Green's function method [35, 36].

## II. EXPERIMENTAL DETAILS

In this study consider two sets of samples. The first one consists of polycrystalline films deposited by magnetron sputtering, used for industrial applications. This procedure for film deposition is considered as a flexible technique, fast and relatively unexpensive. It does not require any particular growth conditions such as ultrahigh vacuum and high temperatures.

However, for a better understanding of the fundamental properties of the system, in particular the study of the electronic band dispersion with angle-resolved photoemission spectroscopy, a high quality of crystalline samples is necessary. For this reason, a second set of samples, composed of crystalline films grown by PLD was prepared.

### 1. Polycrystalline samples

STO films were prepared by RF (13.56 MHz radio frequency) magnetron sputtering using a BOC Edwards TF 600 coating system. The deposition chamber was at a pressure of $2.10^{-6}$ mbar with a STO target (3-inch, purity 99.9 at%) placed inside the chamber. The sample was sputtered at an Argon pressure of $6.10^{-3}$ mbar with an RF power of 400 W onto a Si substrate at room temperature. We produced films of thicknesses of about 200 nm. For Ni-doped STO films deposition, Ni pellets were placed on the erosion zone of the target. The Ni concentration is proportional to the number of Ni pellets used during the growth.

All as-deposited films were found to be amorphous as the growth was performed at room temperature. Such a low temperature is inadequate for STO crystallites formation during the sputtering. All films were annealed in vacuum at 900°C for two hours to induce crystallinity. In order to determine the crystalline structure, were have carried out XRD measurements using an automatic powder diffractometer XPert Pro that employs the Cu $K_\alpha$- line as a source of photons with a wavelength of λ=1.54 Å. XRD patterns were recorded from 20 to 70 degrees in 2θ-scales, where the strongest four STO lines are located.

### 2. Crystalline samples

Pure STO and STO:$Ni_x$ thin films with Ni content of 6% and 12% were grown by PLD on STO substrates with an atomically at (001) surface with the $TiO_2$ termination. In order to reduce charge effects during the photoemission measurements the STO substrates were doped with 5% of Nb. Pressed powder targets, commercially available, with the desired stoichiometry were used (pure STO and $(STO)_{1-x}/(NiO)_x$ targets with x=6%, and 12%). The substrate temperature during the growth was 700°C, and oxygen pressure was $1.10^{-1}$ mbar. Thickness of the deposited films was about 4 nm that is equivalent to 11 unit cells.

The PLD system is furthermore equipped with RHEED and LEED setups to monitor the quality of grown films. An example of the RHEED intensity of the specular reflection during the growth process in shown in Fig. 1(a). The intensity oscillations indicate a layer-by-layer growth mode. There is however a qualitative difference between undoped and Ni-doped films. For the undoped STO films, the oscillations are initially strongly pronounced and then they partially fade after depositing 6 unit cells (t=300 seconds). For the STO:$Ni_x$ films, the oscillations and their amplitudes are more intense and stable throughout the growth.

The post-growth RHEED and LEED patterns are presented in Fig. 1(b) and (c), respectively. For the pure STO films, the diffraction patterns detected in RHEED are typical for a STO(100) surface as well as the pattern observed in LEED. The intensity and contrast of the RHEED patterns increase by increasing the Ni concentration in STO, suggesting that the STO:$Ni_x$ films are better ordered than STO films. The corresponding LEED patterns also show a (1×1) surface reconstruction similar to the STO(100) films, yet the diffraction signal appears to be of higher contrast for the doped samples.

After the growth, our samples were measured in-situ by X-ray photoemission spectroscopy (XPS) with a monochromatic Al $K_\alpha$ x-ray source of 1486.7 eV. X-ray absorption (XAS), resonant photoemission (resPES) and angle-resolved photoemission spectroscopy (ARPES) measurements were performed at PEARL and CASSIOPEE beamlines of the Swiss Light Source (SLS) and SOLEIL Synchrotron, respectively. Samples were transferred to the endstations in a vacuum suitcase ($10^{-10}$ mbar) avoiding exposure to air. Our experiments were done in an ultrahigh vacuum environment of $10^{-10}$ mbar at room temperature and the Fermi level was determined using a gold sample.

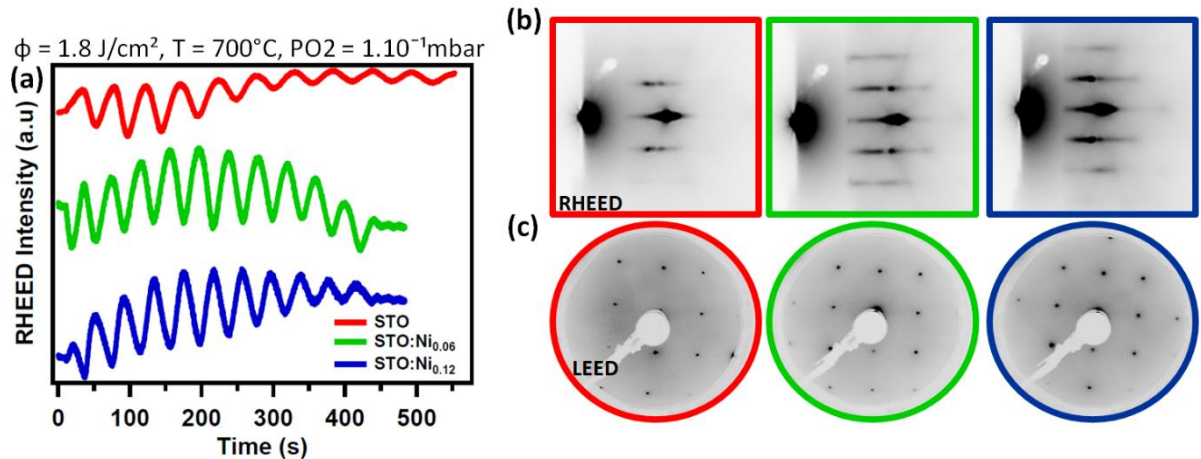

FIG. 1. Structural characterization of thin STO and STO:$Ni_x$ films grown on STO(100) substrate by PLD. (a) Intensity oscillations of the specular RHEED reflection. (b) Corresponding post-growth RHEED patterns of the films. (c) LEED patterns of the undoped and Ni-doped STO films taken with an electron energy of 75 eV. All patterns exhibit a (1×1) reconstruction typical for the cubic perovskite (100) surface.

### III. COMPUTATIONAL DETAILS

All our calculations were performed within the frame-work of the Korringa Khon-Rostoker Green's function method (KKR-GF) based on multiple scattering theory for the calculation of the electronic structure of materials [35, 36]. The potential energy for the STO (100) bulk system with $TiO_2$ termination is induced in a self-consistent way using the atomic sphere approximation (ASA) together with the local density approximation (LDA). The impact of chemical disorder is handled by means of the Coherent Potential $_x$and a new potential is generated. The lattice parameter used in this study is 3.905 Å. The $l_{max}$ value was fixed to 3 (f electrons) for the expansion in spherical harmonics. This will typically improve the calculated value for the band gap as compared to experimental values.

Our ARPES calculations are based on the so-called one step model of photoemission [39] in which experimental parameters such as measurement geometry, photon energy, binding energy and k-vector resolutions, have been taken into account as input in this theoretical investigation.

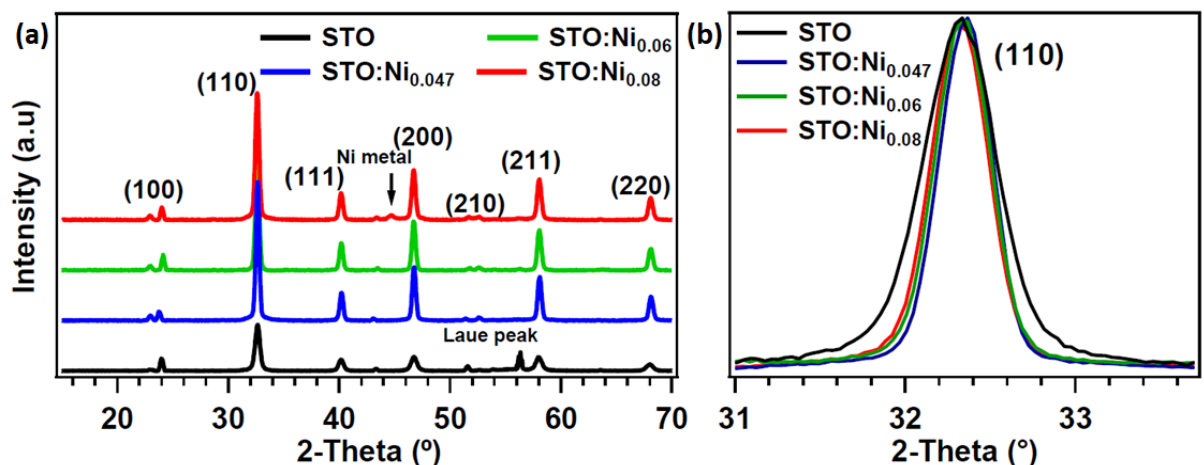

FIG. 2. (a) XRD patterns of annealed pure polycrystalline STO and STO:$Ni_x$ films. (b) Normalised (110) diffraction lines for STO and STO:$Ni_x$ films with (x=0.047, 0.06, and 0.08).

## IV. RESULTS AND DISCUSSION

### A. X-ray Diffraction on Polycrystalline Films

The XRD patterns of pure STO and STO:Ni$_x$ with x=0.047, 0.06, and 0.08 are presented in Fig. 2 (a). All films are showing a polycrystalline structure for a typical cubic STO lattice (ref. JCPDS: 35-0734) composed of crystallites of various sizes and orientations. The additional peak positioned at 44.5° in case of STO:Ni$_{0.08}$ films results from the formation of a Ni metallic phase in the sample (JCPDS: 04-0850).

The first qualitative inspection reveals that the relative intensity of the (110) diffraction lines increases with Ni-doping, indicating that the crystalline quality of the films increases. The normalised (110) diffraction lines are shown in Fig. 2(b).

Crystallite sizes, D$_{XRD}$, listed in in Table I, are calculated from the X-ray line broadening of the (110) plane using Scherrer's equation:

$$D_{XRD} = k\lambda/(\beta cos\theta)$$

where D$_{XRD}$ is the mean size of the crystallite, _ is the X-ray wavelength, k is the shape factor equal to 0.98, θ is the diffraction angle and β is the full width at half maximum (FWHM) [40].

The data show remarkable differences in the width of the lines of the pure and doped films. The diffraction lines of STO:Ni$_x$ films are sharper in comparison to pure STO films by a factor of 1.2. This means that the mean crystallite size increased from 28.2 nm for the undoped films to 34.6 nm after doping, reflecting a better atomic ordering and crystallinity of the films.

In fact, Ni in STO acts as an acceptor and induces the formation of oxygen vacancies that play an important role in structural mobility. If the mobility of oxygen vacancies is high enough, the interaction between the dopant and grain boundaries becomes relatively weak and larger crystallites are formed. Moreover, the difference in size between Ni and Ti can induce a strain field that modifies the grain growth process and may increase the crystallite size. Nevertheless, this observation opens the question on how Ni-doping modifies the electronic properties of STO.

TABLE I. Size of crystallites oriented in the (110) direction for undoped and Ni-doped STO films.

| Sample | 2θ (°) | FWHM (°) | D$_{XRD}$ (nm) |
|---|---|---|---|
| STO | 32.25 | 0.319 | 28.21 |
| STO:Ni$_{0.047}$ | 32.29 | 0.26 | 34.63 |
| STO:Ni$_{0.06}$ | 32.29 | 0.27 | 33.34 |
| STO:Ni$_{0.08}$ | 32.25 | 0.29 | 31.03 |

### B. Photoelectron Spectroscopy Measurements on STO(100) & STO:Ni(100) Films

#### 1. XPS measurements

In order to identify the stoichiometry of the samples, and the chemical environment of Ni impurities, the core level spectra for undoped and STO:Ni$_x$ films were measured by XPS, shown in Fig. 3. The XPS spectra of Sr 3d, Ti 2p and O 1s in STO, STO:Ni$_{0.06}$ and STO:Ni$_{0.12}$ films, displayed in Fig. 3(a-c), respectively, indicate that the individual spectral weights are independent of the presence of Ni. No detectable contamination by other elements, namely carbon, is present.

A comparison of the Ni 2p spectra is presented in Fig. 3(d): In case of STO:Ni$_{0.06}$ and STO:Ni$_{0.12}$, the binding energy positions of the Ni 2p-doublet are found at 855.6 eV (Ni 2p$_{3/2}$) and 873.2 eV (Ni 2p$_{1/2}$) separated by ΔE=17.6 eV. In terms of orbital configuration, these peaks correspond to Ni 2p final state c3d$^9$L. At around 6 eV higher binding energy from each doublet line, satellite features are visible suggesting the oxidised state of Ni atoms in STO films [41] that corresponds to c3d$^8$ and c3d$^{10}$L$^2$ final states. As expected, the spectral weight is increasing by a factor of two for the STO:Ni$_{0.12}$ sample in comparison to STO:Ni$_{0.06}$.

The Ni 2p$_{3/2}$ spectra are fitted with Voigt functions as depicted in Fig. 3(e) and (f). For STO:Ni$_{0.06}$ films, fitting of the Ni 2p$_{3/2}$ line results in two components positioned at 855.8 eV and 857.7 eV (ΔE=1.9 eV) labelled #1 and #2, respectively. The first one is assigned to the oxidation state Ni$^{2+}$, while the second component can have different explanations which we are going to discuss further on. Similarly for STO:Ni$_{0.12}$ films, two components of the Ni 2p$_{3/2}$ peak are fitted. The main Ni 2p line is positioned at 855.4 eV (#1) and the second component (#2) at 857 eV (ΔE=1.6 eV). However, the spectral weight of the two components is not the same for the different samples. In case of STO:Ni$_{0.06}$, the intensity ratio (#1)/(#2) is equal to 2.86, while in STO:Ni$_{0.12}$ the contribution of the component #2 increases significantly and the ratio decreases to 1.12.

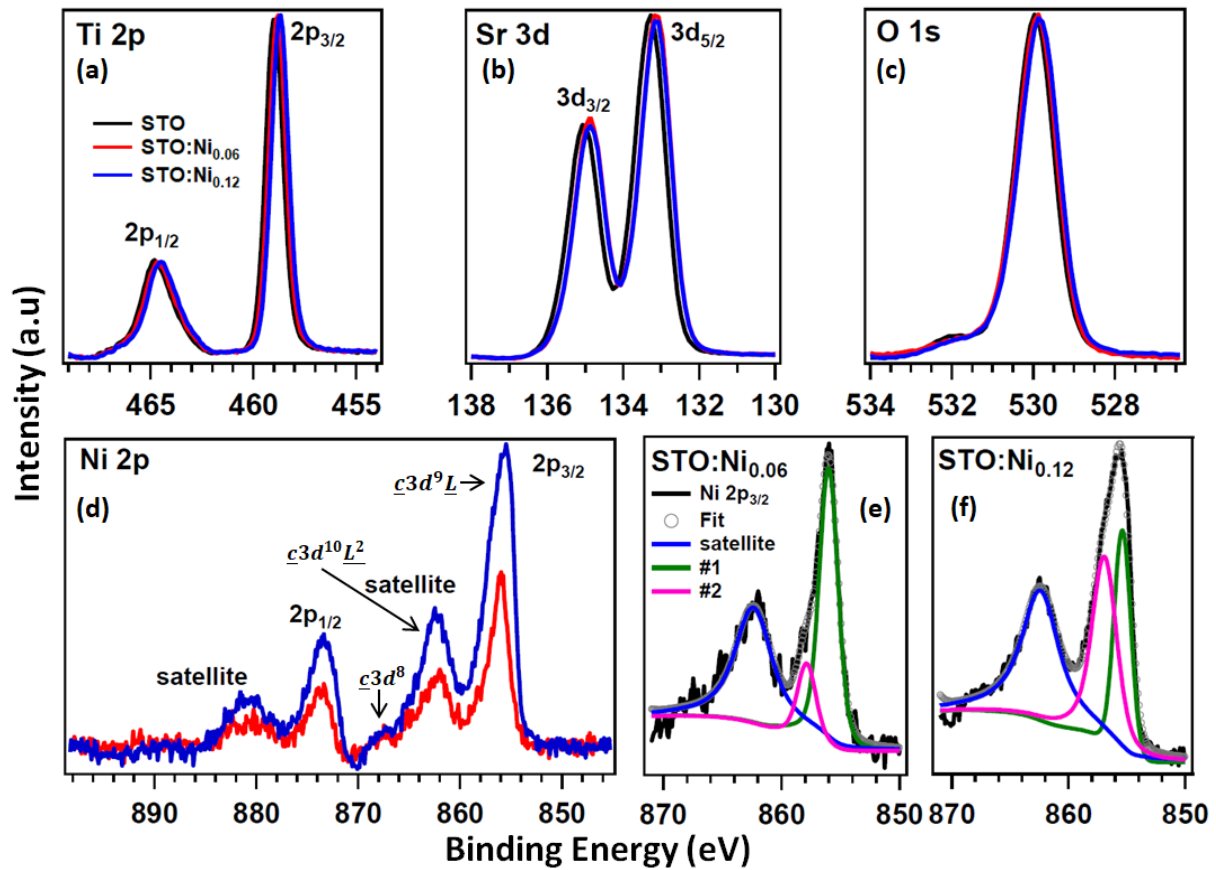

FIG. 3. (a-d) Core level intensity obtained with hv=1486.6 eV at normal emission for STO, STO:Ni$_{0.06}$ and STO:Ni$_{0.12}$ films. All spectra are normalised by the Sr 3d spectral weight. (e-f) Ni 2p$_{3/2}$ spectra of STO:Ni$_{0.06}$ and STO:Ni$_{0.12}$ films, respectively. Voigt functions are used for fitting after Shirley background subtraction.

For the interpretation of the extra component #2 located at the higher binding energy side of the main Ni 2p line we assume that Ni has an octahedral environment with oxygen ions in the films. The formation of component #2 results from a non-local screening process by an electron that does not come from the oxygen orbitals around the Ni atom with the core hole, but from a neighbouring NiO$_6$ unit. Therefore, we define the #1 component as a locally screened core hole $c3d^9L$ and #2 as a non-locally screened core hole $c3d^9L(3d^8L)$. Such an explanation was proposed previously to explain a similar structure of the Ni 2p XPS spectrum in NiO [42, 43]. This hypothesis is verified by the difference seen in the spectral weight of the component #2 as a function of Ni-dopant concentration in STO films, in which its contribution in STO:Ni$_{0.06}$ is relatively weak and becomes more important in STO:Ni$_{0.12}$. According to our previous statement, this is expected for a Ni rich environment, where the probability of having a $d^8L$ states available as next nearest neighbours to a particular Ni ion, that provide the non-local screening, is much higher.

In addition to that, the #2 peak might also have some surface state contribution resulting from pyramidally coordinated Ni atoms located at the surface of the films and overlaps with the non-local screening peak [44, 45].

Angle integrated valence bands (VB) of undoped, STO:Ni$_{0.06}$ and STO:Ni$_{0.12}$ films are displayed in Fig. 4 (a$_1$). The main features observed in all three spectra at binding energies of 3 eV and 8 eV correspond to the O 2p and Ti 3d states. For the doped samples, a formation of in-gap states located at the binding energy of 2-2.5 eV is indicated by arrows.

A more quantitative view of these new states is obtained by subtracting the VB spectrum of the undoped film from the doped ones, see Fig. 4(a$_2$). Clearly, their intensity is in line with the Ni content, indicating that this is a fingerprint of Ni 3d impurity levels in the STO gap.

In order to verify the origin of the electronic bands modification, in Fig.4(b) we present calculated total (TDOS) and partial (PDOS) densities of state of Ni-doped STO and compare them to the undoped STO. Note that the calculated band gap width in the undoped STO is 2 eV, which is lower than the experimental value. This is due to the shortcoming of DFT in describing the exchange-correlation potentials using the LDA approximation. Here, we neglect the band gap error for all systems, assuming that the relative change in the band gap of doped STO with respect to pure STO can provide the effect of Ni-doping.

The PDOS shows that the VB of pure STO is composed mostly of the O 2p states and the conduction band (CB) is dominated by Ti 3d states. For STO:$Ni_{0.06}$, a wide distinct defect peak appears in the middle of the band gap, close to the VB maximum (VBM). As evidenced by the PDOS, it corresponds to Ni 3d states which are hybridized with O 2p states. The intensity of this peak increases by a factor of two in STO:$Ni_{0.12}$. A closer inspection of PDOS unveils the Ni 3d contribution to the whole VB and as well to the CB edge.

The ground state band structures of STO, STO:$Ni_{0.06}$ and STO:$Ni_{0.12}$ are presented in Fig. 4(c). In the undoped STO, which is our reference, the band gap width at the Γ-point is equal to 2 eV. For STO:$Ni_{0.06}$, an atom-like at defect band crosses the Brillouin zone resulting in a considerable reduction of the band gap size. More precisely, the defect band is formed at 0.65 eV above the VBM at 1.35 eV below the conduction band minimum (CBM).

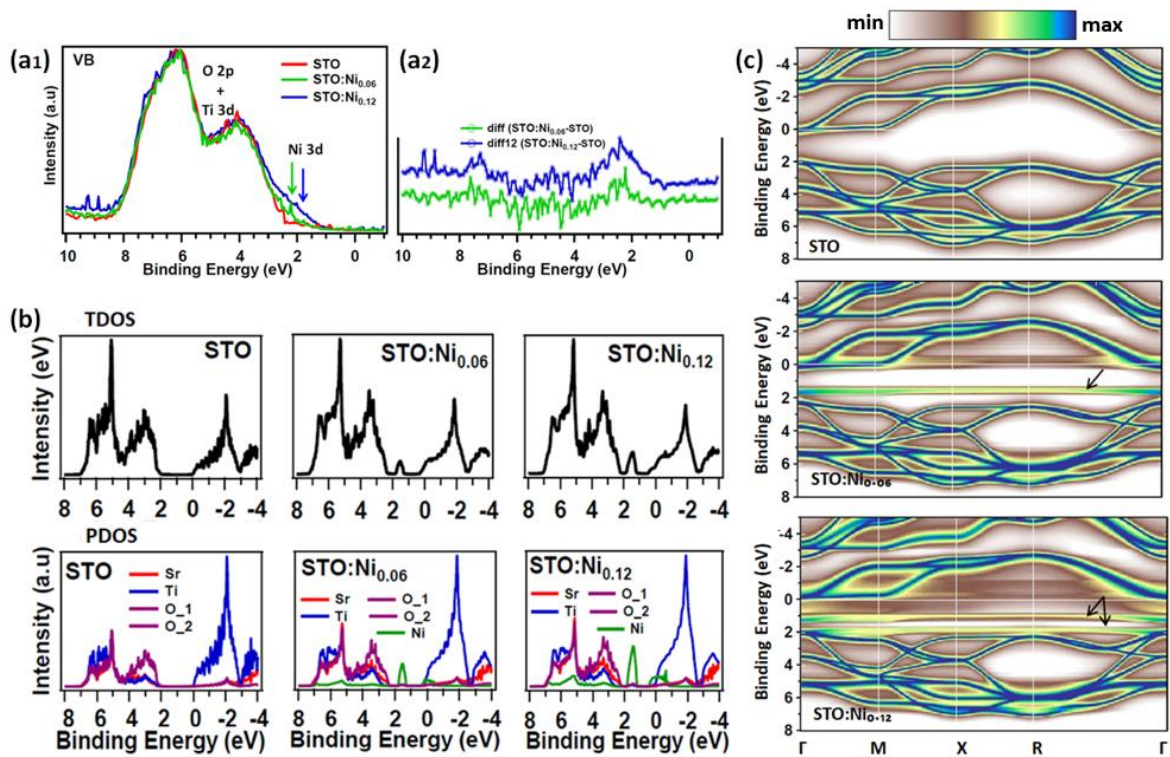

FIG. 4. (a) Angle integrated valence band of STO, STO:$Ni_{0.06}$ and STO:$Ni_{0.12}$ registered with monochromatic hv=1486.6 eV X-rays directly after film growth (samples in-situ). The green and blue arrows assign the formation of in-gap states at the top of the valence band of the doped STO films at around 2-2.5 eV. A quantitative estimation of the new impurity states is given by subtracting from the VB of the doped films with the undoped one. (b) TDOS and PDOS for pure STO, STO:$Ni_{0.06}$ and STO:$Ni_{0.12}$. (c) Calculated ground state band structures (Bloch spectral function) of the corresponding systems. Black arrows are pointed to the Ni 3d levels in the band gap. The Fermi level is set to zero.

By increasing the Ni-doping concentration in STO to 12%, all bands in the VB are shifted by 0.33 eV towards the Fermi level, and two isolated impurity energy levels (IELs) are formed: The first one is located at the VBM at around 2 eV binding energy and the second one is located in the band gap at 0.65 eV above the VBM and at 1.02 eV below the CBM.

In addition to the energy levels formed in the band gap of STO, a new impurity bands are present in the CB. Even though they do not participate in a modification of the band gap, they contribute to the electron density in the VB and/or CB of STO. Moreover, the electronic configuration of Ni is $3d^8 4s^2$, and Ni in the Ti-site is expected to have a 3+ and/or 2+ state ($3d^5 4s^2$ and/or $3d^6 4s^2$) unlike the oxidation state of Ti atoms (4+). This fact leads to the increase of hole concentration and/or oxygen vacancies, so Ni atoms are acceptor impurities [46].

### 2. XAS and resPES measurements

We used resPES to investigate the correlated electronic structure of the 3d Ni states in STO. At energies close to the Ni L-edge absorption threshold, an interference between the direct photoemission channel and the dipole transition of core electron to an unoccupied state (2p→3d) occurs, giving rise to strong variations of the photoemission cross section [47{50].

Prior to resPES measurements we recorded $L_{2,3}$-edge XAS spectra from STO:$Ni_{0.06}$ and STO:$Ni_{0.12}$ films. We note that for both systems, the Ni XAS spectra have overall similar features, consequently only the STO:$Ni_{0.12}$ spectrum is presented in Fig. 5(a). The main peaks of both $L_3$ (853 eV) and $L_2$ (870.2 eV) edges are labelled #a. Two less intense features, lying at approximately 2 eV and 6 eV from the mains peaks, are labelled #b and #c, respectively. The structure #b is due to multiplet splitting of the $2p^5 3d^9$ state, resulting from the Coulomb interaction, exchange interaction and crystal field effects, and #c is attributed to the satellite structure [50].

A zoom on the $L_3$-edge for STO:$Ni_{0.06}$ and STO:$Ni_{0.06}$ films is presented in Fig. 5(b). The main line (#a) exhibits a broadening with increasing Ni concentration due to an additional feature, indicated by arrows at the higher energy side. This broadening results mainly from a non-local screening process that involves the $Ni^{2+}$ and the ligand states [42].

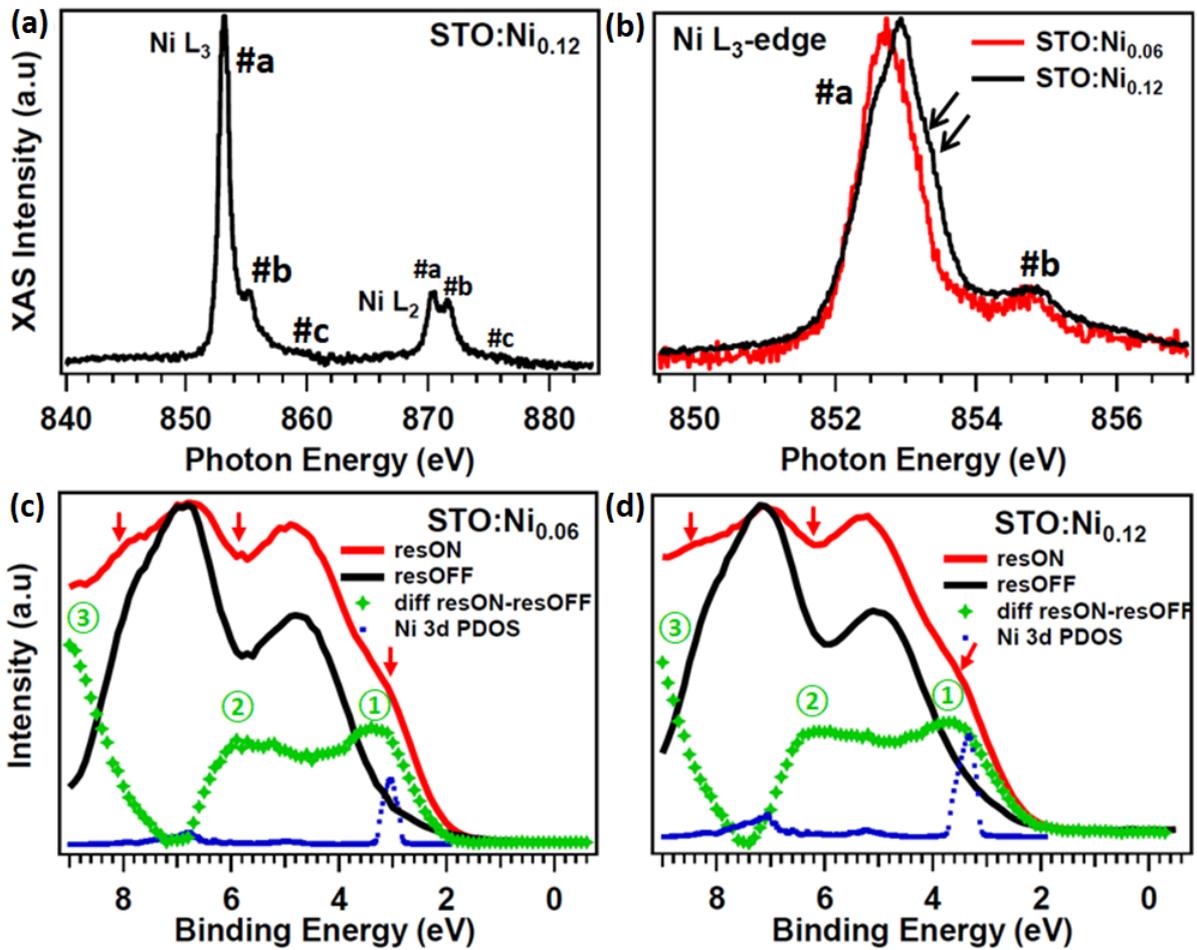

FIG. 5. (a) Ni $L_{2,3}$ absorption spectra for STO:$Ni_{0.12}$. (b) Comparison of the $L_3$ absorption spectra for STO:$Ni_{0.06}$ and STO:$Ni_{0.12}$ films. Resonant photoemission measurements for the valence band of (c) STO:$Ni_{0.06}$ and (d) STO:$Ni_{0.12}$ films registered with two photon energies. resON: Photon energy around the TM $L_3$-edge absorption (hv=853.2 eV). resOFF: Photon energy below threshold (hv=840 eV). The difference of the valence bands resON-resOFF, as well as the 3d-TM PDOS of STO:TM systems are plotted.

In order to determine the resonance enhancement of Ni 3d related parts of the VB, spectra are taken at two photon energies: one at the maximum of the 3d Ni absorption L-edge (resON), and the second below threshold (resOFF). The difference (resON resOFF) presented in Fig. 5(c,d) gives the resonance contribution where we identified three major features labelled 1, 2 and 3. The maxima of respective features lye at binding energies 3.5 eV, 5.5 eV and 9 eV. The calculated Ni 3d PDOS (blue line) shows one main peak that matches only with feature 1. Feature 2 is a resonance of the multiplet structure, reflecting the strongly-correlated nature of Ni-ions with the crystal field. Feature 3 in mainly related to an incoherent Auger decay not participating in the resonance photoemission process and resulting from the satellite located at 6 eV higher binding energy from Ni L3 [51-53].

We note that the LDA approximation, used in our calculations, does not include correlation effects. Consequently, the structures 2 and 3 which involve correlations, cannot be obtained.

### 3. ARPES study

$k_z$ dependent ARPES calculations for bulk STO(001) and STO:Ni$_{0.12}$(001) together with the experimental data for STO:Ni$_{0.12}$(001), all plotted at the binding energy of 4 eV, are shown in Fig. 6(a - c). The first Brillouin zone of STO and the experimental geometry are shown in Fig. 6(d and e). The conversion from hν to $k_z$ is carried out within the free electron final state approximation, with the inner potential of 14.65 eV retrieved from the periodicity observed in the measured spectra. The high symmetry points at $k_x$=0 Å$^{-1}$, Γ and Z, and at $k_x$=0.8 Å$^{-1}$, X and A, as labelled in Fig. 6(d).

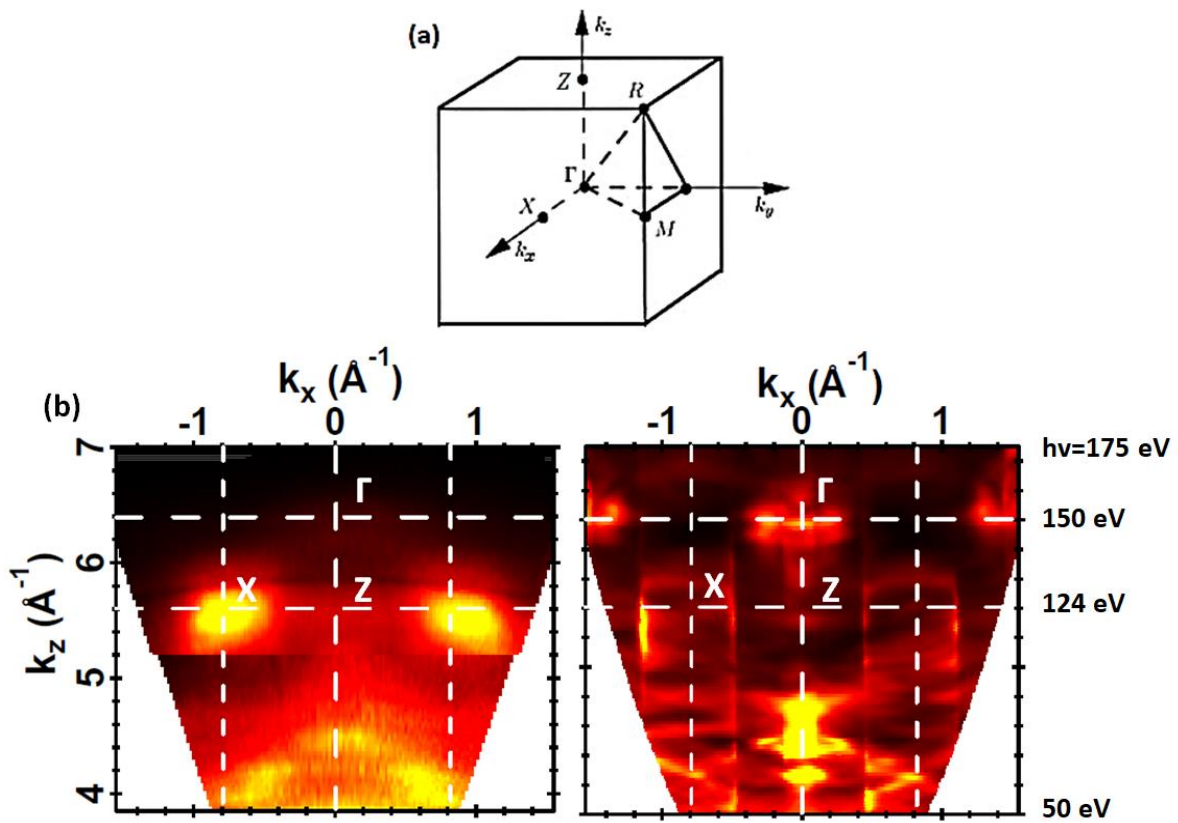

FIG. 6. (a) Dispersion with $k_z$ at BE=4 eV below Fermi energy from hν=50 eV to hν=175 eV showing the high symmetry points. The computational data from the one-step model of photoemission for bulk STO and STO:Ni$_{0.12}$ are displayed in (a) and (b), respectively. (c) Experimental data for STO:Ni$_{0.12}$ films. (d) First Brillouin zone of STO. (e) Measurement geometry.

At a first glimpse, the main features in both, bulk STO(001) and STO:Ni$_{0.12}$(001) calculated spectra, are very similar, see Fig. 6(a,b). The principal difference is the intensity embodied in the background all over the $k_x$ and $k_z$ ranges in case of STO:Ni$_{0.12}$. This intensity is purely related to the localised Ni impurity bands formed at the binding energy level, and confirms that these states are located not only around a specific symmetry point or direction, but they are homogeneously distributed in the whole volume of the Brillouin zone for a given binding energy. These dissimilarities result from a slight upward binding energy shift of the bands for STO:Ni$_{0.12}$ compared to STO, as discussed in the next paragraph. The experimental data displayed in Fig. 6(c) for STO:Ni$_{0.12}$(001) films are in good agreement with the theoretical data except for the intensity of the symmetry points at higher photon energy. The line profile plotted for hν=150 eV confirms the position of the Γ and X-points. The Z-point is detected at hν=124 eV. These values correspond to the third Brillouin zone along the $k_z$-axis.

Experimental band dispersion and corresponding ARPES calculations (one-step model of photoemission) for the STO and STO:Ni$_{0.12}$ along the ZA direction are displayed in Fig. 7. All data are registered with linearly polarized light of 200 eV for a wide binding energy range (10 eV below Fermi level). Note that for one-step calculations, a correction of Fermi level position in applied so that it is compatible with the experimental value. Let us first discuss the experimental data. In the case of pure STO(001), the most significant feature is the dispersion of the O 2p bands partially hybridized with Ti 3d bands between 4.5 eV and 10 eV below Fermi energy.

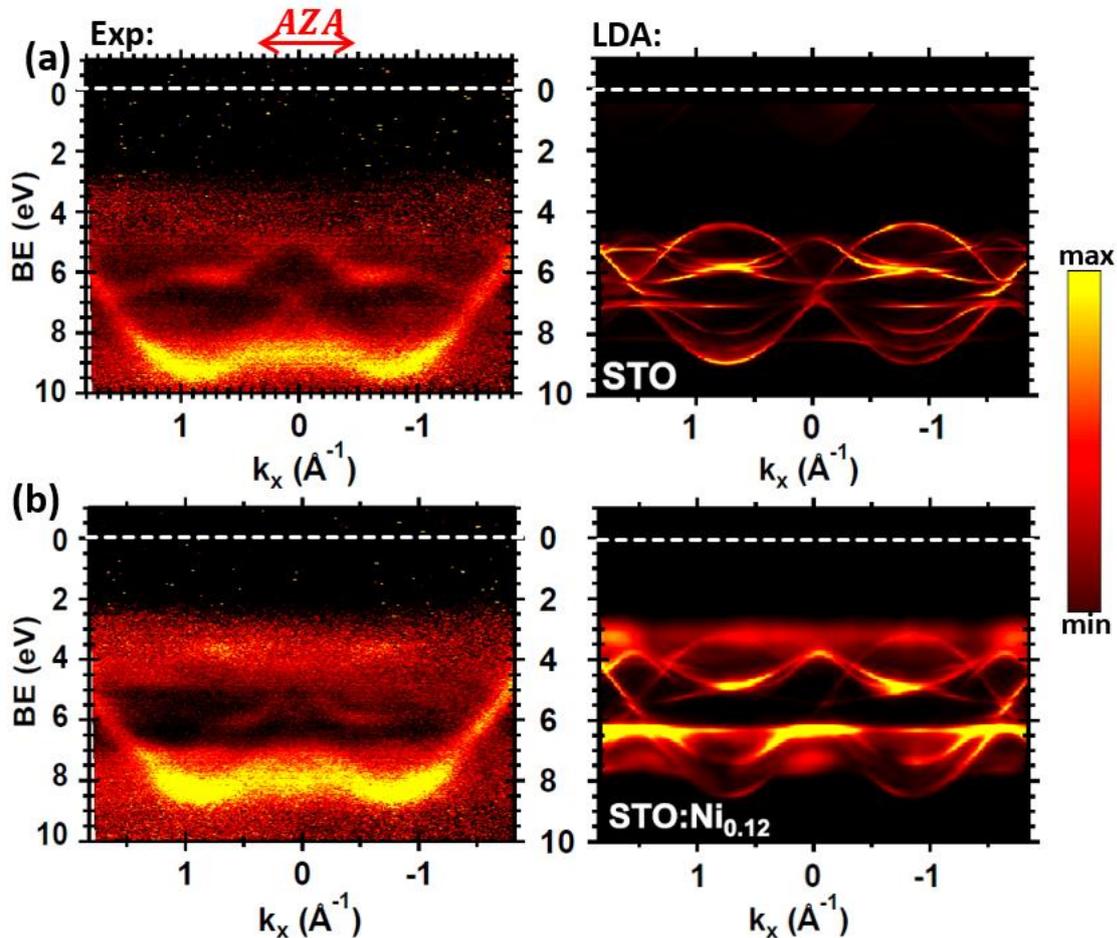

FIG. 7. Experimental band dispersion along the $k_x$ direction for $k_y=0$ Å$^{-1}$, and corresponding ARPES calculations based on one-step model of photoemission, respectively, for (a) STO(001) and (b) STO:Ni$_{0.12}$(001). Data shows the bulk bands dispersion and the localized impurity bands at the top of the valence band for the 3d-TM doped STO films. All data are acquired with linearly polarized light of 200 eV. The measurements were performed at room temperature. Fermi level is set to zero.

For clarity, only data of STO:Ni$_{0.12}$(001) films are presented (in Fig. 7(b)). Note that no major differences between STO:Ni$_{0.12}$(001) and STO:Ni$_{0.06}$(001) are detected but only some minor intensity variations in particular when it concerns the impurity states.

In case of Ni-doped STO films, the main bulk band between 7 eV and 9 eV is shifted to lower binding energy level by 1 eV in comparison to pure STO films. The upper bands between 4 eV and 6 eV are shifted by 1.3 eV to lower binding energy.

The ARPES calculation for STO shows mainly the O 2p bands partially hybridised with Ti 3d bands (see right hand plot in Fig. 7(a)). Yet, the bands width is slightly underestimated with respect to experimental data. As already stressed, this disagreement reflects the limitations of DFT with LDA especially when it deals with strongly correlated systems. Despite the shifted position of Fermi level, the calculated band dispersion of STO:Ni$_{0.12}$ reproduces quite well the measured one, notably the localised Ni 3d energy levels formed at the top of the VB (see right hand plot in Fig. 7(b)).

## V. CONCLUSIONS

STO doped with Ni are investigated. The samples were synthesised as thin films of a few nm thickness. A variety of photoelectron spectroscopy techniques such as XPS, XAS, resPES and ARPES were applied to achieve understanding of the complicated many body physics. In addition to experimental investigations, first-principle calculations within the KKR-GF method were performed.

Using XRD, we showed that for polycrystalline Ni-doped STO films, grown by magnetron sputtering deposition, the crystallite size increased considerably and the films micro-structure improved compared to the non-doped films.

In case of STO:Ni(001) films, grown by PLD, the analysis of the Ni 2p core level spectra confirmed the oxidised form of the impurity atoms in STO, and their location in an octahedral environment. From first principle calculations for ground states and from the resPES measurements with photon energies around the 3d Ni L-edge, we could define the distribution of the 3d impurity energy levels in the VB/CB and the band gap of STO. The localised defect bands in the band gap at the VBM will lead to an improvement of visible light absorption. The Ni-resonant measurements revealed the strongly correlated nature of Ni 3d electrons in STO, where multiplet and satellite structures have an important contribution to the Ni L-edge resonance. The origin of these structures results mainly from strong Coulomb interaction, exchange interaction and crystal field effects. ARPES data indicate that for the doped films, the bands are shifted towards Fermi level, and the localised 3d impurity bands were detected at 2.5 eV binding energy in the band gap of STO. One-step model calculations (LDA) reproduce nicely the band dispersion, however we observe discrepancies in both, the Fermi level position and in the bands width. This can be explained by the fact that the calculations do not include all correlation effects and they underestimate the Coulomb potential.

## VI. ACKNOWLEDGMENTS

We gratefully acknowledge interesting discussions with V. Strocov. This work was mainly supported by the CEDAMNF project which is financed by the Ministry of Education, Youth and Sports of Czech Republic, Project No.CZ.02.1.01/0.0/0.0/15.003/0000358.